# Feasibility of optical probing of relativistic plasma singularities

Timur Zh. Esirkepov[1], Jie Mu[2], Yanjun Gu[2], Tae Moon Jeong[2], Petr Valenta[2], Ondrej Klimo[2], James K. Koga[1], Masaki Kando[1], David Neely[3,4], Georg Korn[2], Sergei V. Bulanov[1,2], Alexander S. Pirozhkov[1]

[1]Kansai Photon Science Institute, National Institutes for Quantum and Radiological Science and Technology, 8-1-7 Umemidai, Kizugawa-city, Kyoto 619-0215, Japan

[2]Institute of Physics of the ASCR, ELI Beamlines Project, Na Slovance 2, 18221 Prague, Czech Republic

[3]Central Laser Facility, Rutherford Appleton Laboratory, STFC, Chilton, Didcot, Oxon OX11 0QX, UK

[4]Department of Physics, SUPA, University of Strathclyde, Glasgow G4 0NG, UK

**Abstract.** Singularities in multi-stream flows of relativistic plasmas can efficiently produce coherent high-frequency radiation, as exemplified in the concepts of the Relativistic Flying Mirror [S. V. Bulanov, *et al.*, *Phys. Rev. Lett*. **91**, 085001 (2003)] and Burst Intensification by Singularity Emitting Radiation (BISER) [Pirozhkov, *et al.*, *Scientific Reports* **7**, 17968 (2017)]. Direct observation of these singularities is challenging due to their extreme sharpness (tens of nanometers), relativistic velocity, and transient non-local nature. We propose to use an ultrafast (a few light cycles) optical probe for identifying relativistic plasma singularities. Our Particle-in-Cell (PIC) simulations show that this diagnostic is feasible.

## Introduction

Singularities easily form in multi-stream flows; they appear as crests or surges with respect to some physical parameter, which is usually continuous, e.g., density, pressure, etc. If these crests and surges correspond to structurally stable singularities, they are inevitable and robust with respect to modulations, as explained by catastrophe theory [1, 2]. In relativistic underdense plasmas driven by intense femtosecond lasers, two particularly interesting examples of the cusp singularities appear in electron density: one in a breaking wake wave [3] and another at the joining of the cavity wall and the bow wave [4]. In order to create these singularities, the driving laser should be sufficiently intense, with the dimensionless amplitude of the order of 1 or greater, $a_0 = eE_0/m_ec\omega_0 = (I_0/I_R)^{1/2} \gtrsim 1$, and short, with the length comparable to or shorter than the Langmuir wavelength, $c\tau \lesssim \lambda_{pe}$. Here $c$ is the speed of light in vacuum, $\tau$ is the pulse duration; $E_0$, $I_0$, $\lambda_0$ and $\omega_0$ are the laser electric field, irradiance, wavelength and angular frequency, respectively; $e$ and $m_e$ are the electron charge and mass, respectively; $I_R = \pi c^5 m_e^2/2e^2\lambda_0^2 \approx 1.37\times10^{18}$ W/cm$^2\times(\lambda_0[\mu m])^{-2}$ for a linearly polarized field; $\lambda_0[\mu m]$ denotes the wavelength in micrometres. We consider underdense plasma, with the initial electron density much less than the critical density, $n_e << n_{cr}$, where $n_{cr} = m_e\omega_0^2/4\pi e^2 \approx 1.1\times10^{21}$ cm$^{-3}/(\lambda_0[\mu m])^2$.

The first mentioned above cusp singularity appears when a wake wave driven by a relativistically intense laser pulse breaks. Under optimum conditions, this cusp can take the form of a dense, thin shell moving behind the driving laser pulse with relativistic velocity. Due to the high density and sharpness, this shell can act as a Relativistic Flying Mirror, reflecting a counter-propagating laser pulse [5-8]. The frequency of the reflected light is upshifted and duration shortened due to the double Doppler effect, so that a high-frequency coherent pulse is produced; in addition, the shell can have a concave, nearly parabolic, shape [6], which focuses the reflected radiation to a tiny spot, promising record intensities [6, 9] beyond capabilities of directly focused lasers. The Relativistic Flying Mirror was demonstrated experimentally in [10-17], where the frequency upshift was sufficient to convert a near infra-red laser radiation ($\lambda_0 \approx 0.8$ μm) to coherent soft x-rays, down to $\lambda_x = 7$ nm.

The second mentioned cusp singularity appears when the bow wave detaches from the cavity wall, which happens when the laser spot is sufficiently narrow, $d < d_{BW} = 2\lambda_0(a_0 n_{cr}/n_e)^{1/2}/\pi$ [4]. This can be achieved, for example, due to relativistic self-focusing [18-20], where the spot size under stationary conditions can be estimated as [21] $d_0 = \lambda_0(a_0 n_{cr}/n_e)^{1/2}/\pi$ and the dimensionless amplitude as $a_0 = (8\pi P_0 n_e/P_c n_{cr})^{1/3}$, where the laser power $P_0$ exceeds the threshold $P_{SF0} = P_c (n_{cr}/n_e)$ and $P_c = 2m_e^2 c^5/e^2 \approx$ 0.017 TW [20]. This cusp singularity is situated near the laser pulse head, and is strongly driven by the laser field. The accelerated motion of the electrons in the singularity leads to high-frequency radiation, while the singularity sharpness ensures constructive interference, producing a bright coherent x-ray pulse [17, 22-27] termed Burst Intensification by Singularity Emitting Radiation (BISER) [28]. BISER has an unprecedentedly small source size: sub-μm measured directly (limited by the resolution of employed x-ray optics) and down to 10 nm predicted by the PIC simulations [28]. Together with the attosecond duration (predicted by PIC) and large photon number (measured $10^{10}$ photons with energy from 60 to 100 eV), this gives a peak spectral brightness of up to $10^{27}$ photons/mm$^2$ mrad$^2$ s 0.1% BW, which is among the highest amongst laser-based sources.

Direct experimental observation of these singularities is challenging due to their small size, relativistic velocity, and transient non-local nature. Here we propose to use an ultrashort-duration (few-cycle) optical probe to detect these singularities, the Schlieren imaging. We specify the necessary probe pulse parameters and present the results of Particle-In-Cell (PIC) simulations to demonstrate the diagnostic feasibility.

The use of ultrashort optical probe pulses has been previously demonstrated for visualizing wake waves in laser plasma [29, 30]. Our study shows that similar technique can be used for visualizing singularities in laser plasma: the probe pulse undergoes strong diffraction on the moving singularities producing spherical waves whose frequency depends on angle due to the relativistic effects. The corresponding upshifts or downshift of the probe frequency can be easily detected at angles viewable with Schlieren imaging. We also show that our method reveals the Lampa-Penrose-Terrell effect.

### Schlieren imaging scheme

A possible experimental setup is shown in Fig. 1. An intense driver laser pulse (further – the driver) with the dimensionless amplitude of $a_0 > 1$ is focused onto a supersonic gas jet. Since the magnitude of the driver field well exceeds that of the intra-atomic field, the gas becomes ionized with a time period shorter than the laser cycle. An underdense plasma is created. The driver power is greater than the threshold of relativistic self-focusing. The self-focusing driver excites a wake wave and bow wave, as seen below in the simulations. As described in the Introduction, the cusps formed near the driver head emit radiation, BISER. It can be seen by an x-ray spectrometer placed within a relatively small angle about the direction of the driver propagation.

A weak probe laser pulse (further – the probe) is irradiated onto the plasma in the transverse direction, perpendicular to the driver axis. It propagates through the channel created by the driver, then goes into a special optical system which makes a Schlieren image as shown in Fig. 2 (a). The probe is diffracted on singularities of electron density and electron current density, formed in the plasma by the driver. These singularities are strongly localized, thus they easily break the approximation of geometric optics. As seen in Fig. 2 (a), a singularity produces a spherical wave, which is imaged by the lens as a point-like source. In this way the singularities can be identified using Schlieren imaging. We note that the probe pulse length must be shorter than the longitudinal and *transverse* distances between the singularities, $c\tau < d_0$, otherwise the spherical waves appear with their centers significantly drifting with time, so that different singularities become indiscernible due to the longitudinal and transverse motion blurs.

We can roughly estimate the cusp singularity effect on the probe phase by assuming a stationary cusp density distribution $n_e = n_0(y/\lambda)^{-2/3}$ [4], Fig. 2(b), and integrating the refractive index difference: $\Delta\phi = (2\pi/\lambda)\int(n-1)dy$, where the refractive index is $n = (1-n_e/n_{cr})^{1/2}$, $\lambda$ is the probe wavelength, and $y$ is along the probe propagation direction. The integration from 0 to $y_{max} = M\lambda$, where $M$ is the dimensionless length, gives the real part $\text{Re}[\Delta\phi] = 2\pi((M^{2/3} - n_0/n_{cr})^{3/2} - M)$, which slowly diverges as $\text{Re}[\Delta\phi] \to -3\pi M^{1/3} n_0/n_{cr}$ at large $M$. We note that in 3D the density starts to decrease quickly at $y > d_0$, so for estimating we can use $M = d_0/\lambda$. The imaginary part is $\text{Im}[\Delta\phi] = 2\pi(n_0/n_{cr})^{3/2}$, corresponding to the transmission intensity of $\exp(-2\pi(n_0/n_{cr})^{3/2})$. For the cusp parameters as in [31] ($n_0 = 0.023 n_{cr}$) this gives the real part of $\text{Re}[\Delta\phi] = -0.46$ rad, and the imaginary part of $\text{Im}[\Delta\phi] = 0.02$ rad, corresponding to the beam transmission of 0.96, i.e. 4% attenuation. Such a phase shift and attenuation are well within measurable ranges.

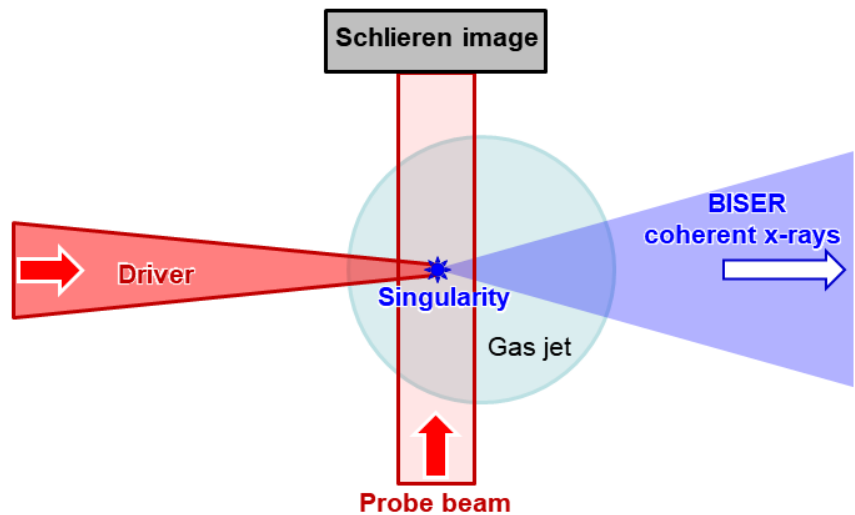

Fig. 1. The proposed experiment setup. The driver laser pulse is focused onto a gas jet. The resulting BISER is detected in the direction of the driver. The singularities are imaged by the probe ultrashort laser pulse with the same wavelength but different polarization.

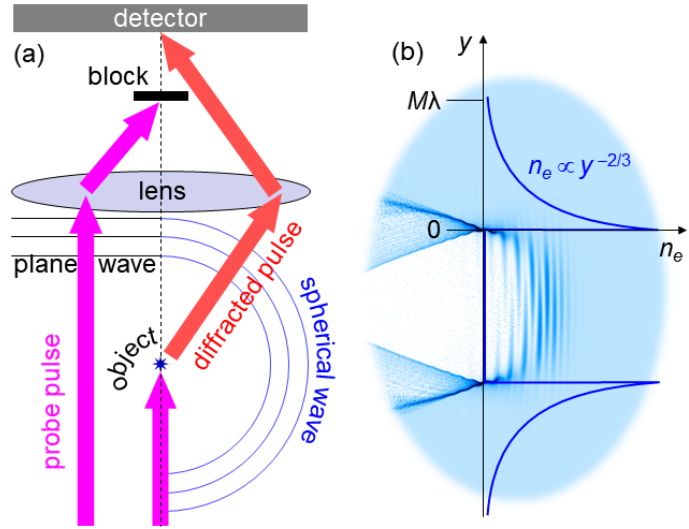

Fig. 2. (a) The principal scheme of the Schlieren imaging. On the left: the rays and wave fronts of the probe pulse; after transmission through the lens the probe pulse is blocked by an opaque filter. On the right: the rays and wave fronts of the probe pulse diffracted at a point object; after transmission through the lens the diffracted pulse is focused onto the detector, forming the image of the object. The actual configuration has rotational symmetry with respect to the axis denoted by the dashed line. (b) The scheme for the refractive index integration (see the text).

## Simulation setup

We have performed two-dimensional (2D) PIC simulations using the REMP code [32]. The setup is shown in Fig. 3. The driver laser pulse propagates in the direction of $x$-axis (in Fig. 3 – from the left to the right); it is $p$-polarized, i. e. in the direction of $y$-axis, in the plane of the simulation box; initially its electromagnetic field is given by $E_x$, $E_y$, $B_z$ components while other components are zero. The driver has a gaussian shape with the full width at half maximum (FWHM) of 5 $\lambda_0$ and the focal spot size of 5 $\lambda_0$. Here $\lambda_0$ is the driver wavelength. The driver amplitude is $a_0 = 6.6$; its assumed focal plane is inside plasma at the distance of 40 $\lambda_0$ from the vacuum-plasma interface, where the driver enters plasma ("assumed", because the value is given for the case of the driver propagation in vacuum).

The plasma slab is rectangular and has the sizes of 160 $\lambda_0$ and 80 $\lambda_0$ in the direction of the $x$ and $y$ axes, respectively. The electron density is $n_e = 0.01\ n_{cr}$.

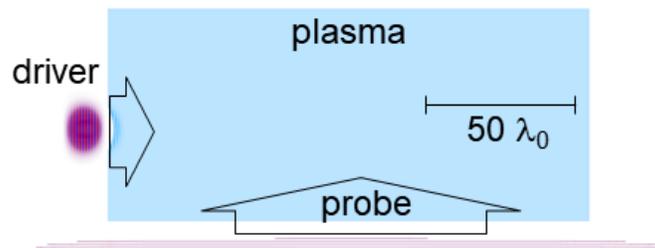

Fig. 3. The PIC simulation setup.

The probe laser pulse propagates in the direction of the *y*-axis (in Fig. 3 – from the bottom to the top); it is *s*-polarized, i. e. in the direction of the *z*-axis, perpendicular to the plane of the simulation box. The difference in the driver and probe polarization helps to distinguish the evolution of the driver and probe. As is well known, the interaction of a *p*-polarized electromagnetic wave with plasmas has special symmetry in a 2D configuration: if only three components of the electromagnetic field, $E_x$, $E_y$, $B_z$, and only two components of the plasma particles momentum, $p_x$, $p_y$, were initially non-zero, then all other components will be zero forever. Thus, in the interaction of a *p*-polarized driver with initially calm plasma, the components $E_z$, $B_x$, $B_y$ can acquire non-zero values only in the presence of the s-polarized probe. The probe has Gaussian shape with the FWHM length of 3 $\lambda_0$ and the focal spot size of 160/3 $\lambda_0$. Its amplitude is $a_0$ = 0.01; its assumed focal plane is at the bottom plasma-vacuum interface (for such a wide probe, its Rayleigh length is much greater than the transverse extent of the plasma slab). The probe is sent into the plasma 50 laser cycles later than the driver, so that the probe propagates through the wake wave when the driver is approximately at the center of the plasma slab. The mesh size is $\lambda_0/16$ in both spatial directions, which is sufficient to see the second harmonic. The time step is 0.999 of the threshold corresponding to the Courant-Friedrichs-Lewy condition. The number of quasi-particles per cell is 1; the plasma consists of electrons with a neutralizing background of immobile ions.

## Simulation results

The results of simulations are presented in Figs. 4-6. In the figures, the time unit is the laser cycle. The driver and probe pulses enter the plasma approximately at *t*=20 and *t*=70, respectively. In the figures the probe pulse is represented in two ways simultaneously. The nearly horizontal red curves are the contours of the $E_z$ field component of the probe corresponding to the value of $E_z$=0.001 (in terms of dimensionless amplitude). The white-red color-scale shows the spatial distribution of the magnitude of the difference between the values of the $E_z$ field component of the probe calculated in two simulations: in one simulation both the driver and the probe pulse are present, in another simulation only the probe pulse is present. This method allows to getting rid of physical, but undesirable, reflection of the probe from the plasma slab boundaries. In addition, the Fourier filter was applied to remove the wave propagating at 90°, similarly to the blocking technique in the Schlieren imaging.

In Fig. 4, one can easily see distortions in the probe transmitted through the wake wave. These distortions are "enriched" by spherical fronts which originate at various singularities appearing in the wake wave. We can identify the singularities formed due to longitudinal and transverse wave breaks. The most prominent diffraction, however, occurs at the cusps near the head of the driver pulse. The corresponding spherical wave fronts can be collected by the lens and form the Schlieren image of the cusps, as in Fig. 2. We note that the cusps move with the group velocity of the driver pulse, which is close to the speed of light in vacuum, *c*. Therefore, the image of the two cusps will be such as if the cusps were slightly rotated with respect to each other, a manifestation of the Lampa-Penrose-Terrell effect. This can be seen in Fig. 4, right column. If the cusps were at rest, the two spherical waves from them would have their centers exactly aligned along the probe pulse propagation, thus we would see just one diffracting object. However, due to the fast motion of the cusps, the probe pulse meets them at different horizontal position. This results in the apparent rotation of the pair of cusps as a whole, so that we will see two separated diffracting objects. For this, the probe pulse must be short enough, as stated above.

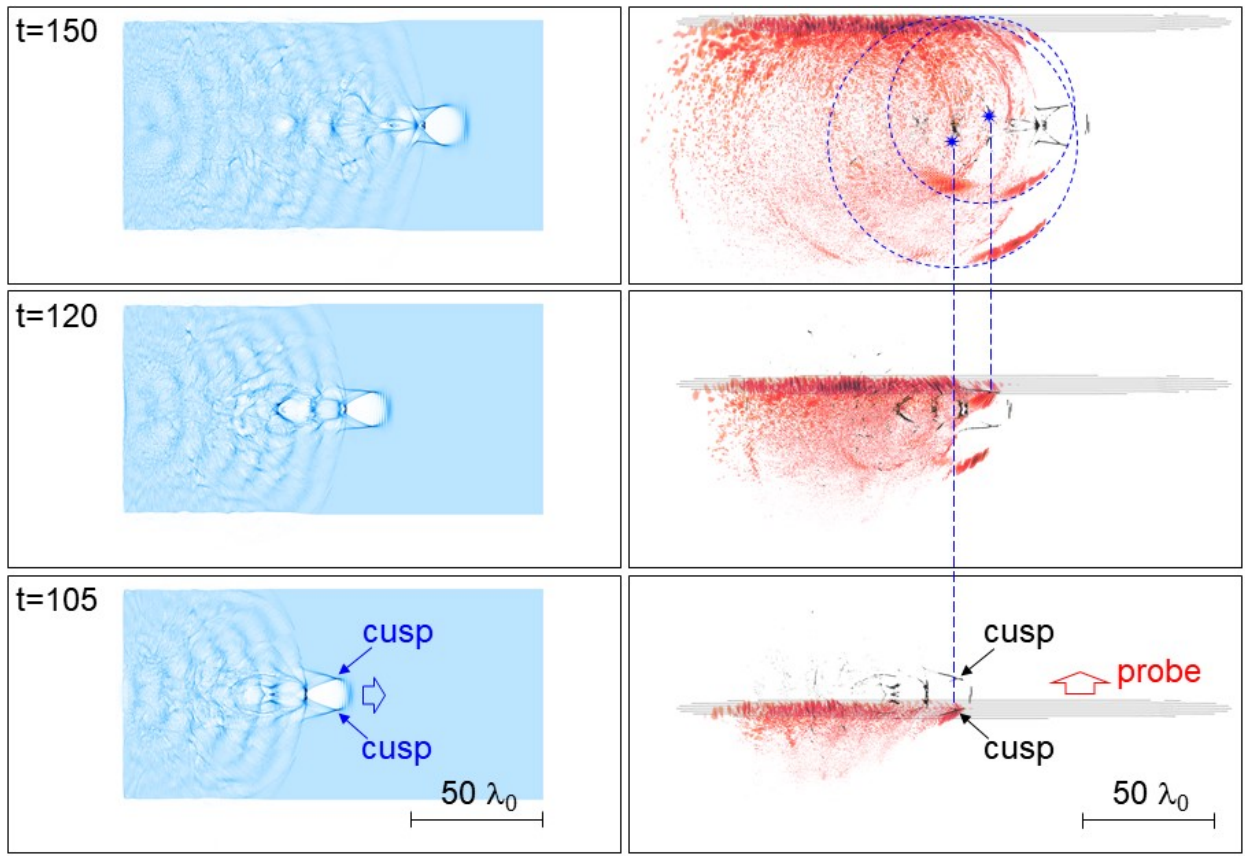

Fig. 4. On the left: electron density at different moments of time (the time unit is the laser cycle). The driver pulse can be traced by density modulations at the front of the first cavity of the wake wave. On the right: black curves are for the electron density (at the levels of density 3 and 4 times greater than the initial density); nearly horizontal red curves are for the electric field of the probe; the white-red color-scale is for the absolute difference of the $E_z$ field component between two simulations, one with the driver and probe and another with the probe only; the waves going at 90° are filtered out (similarly to the Schlieren imaging) using a Fourier filter.

Figs. 5 and 6 show the probe pulse propagation through the wake wave, so one can see were the probe diffraction occurs and how the spherical waves are formed from individual acts of diffraction. In addition, the spatial spectrum of the $E_z$ field component of the probe is shown at different times. The circle $k_x^2+k_y^2=1$ corresponds to waves with the wavelength of $\lambda_0$ going to all possible directions. This circle appears very early due to diffraction of the probe on density modulations at the plasma-vacuum interface.

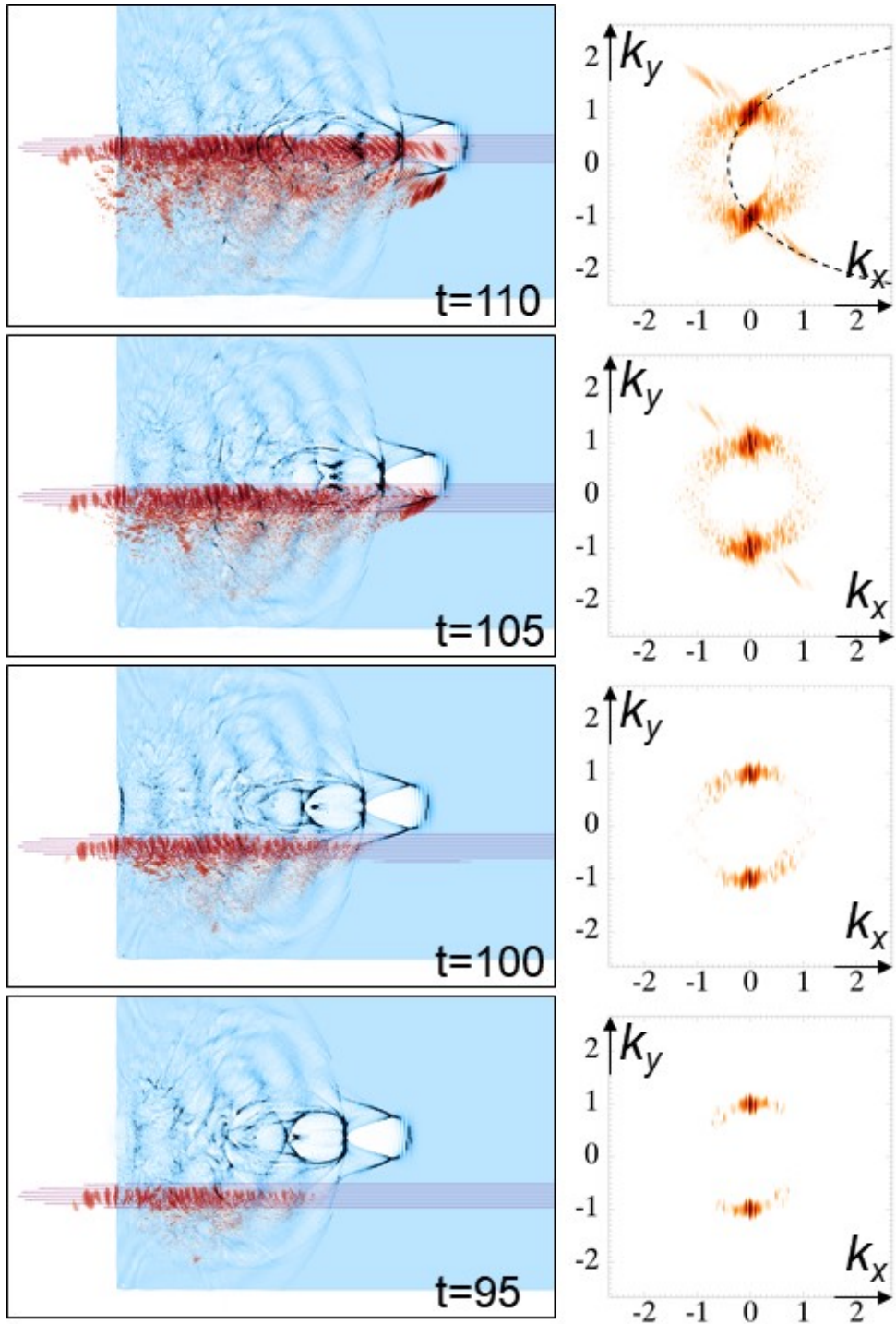

Fig. 5. On the left: the probe propagation through the wake wave; the curves and color-scales are the same as in Fig. 4. On the right: the magnitude of the spatial spectrum of the $E_z$ field component of the probe at the same instances of time as the spatial distributions on the left. The spectral magnitude has maxima at two darkest spots at $k_x=0$ and $k_y=\pm 1$, which correspond to initial probe beam direction and laser wavelength $\lambda_0$. The appearance of high-frequency "protrusions" out of the circle $k_x^2+k_y^2=1$ correlates with the onsets of diffraction from the cusps. The "protrusions" are the portions of an ellipse.

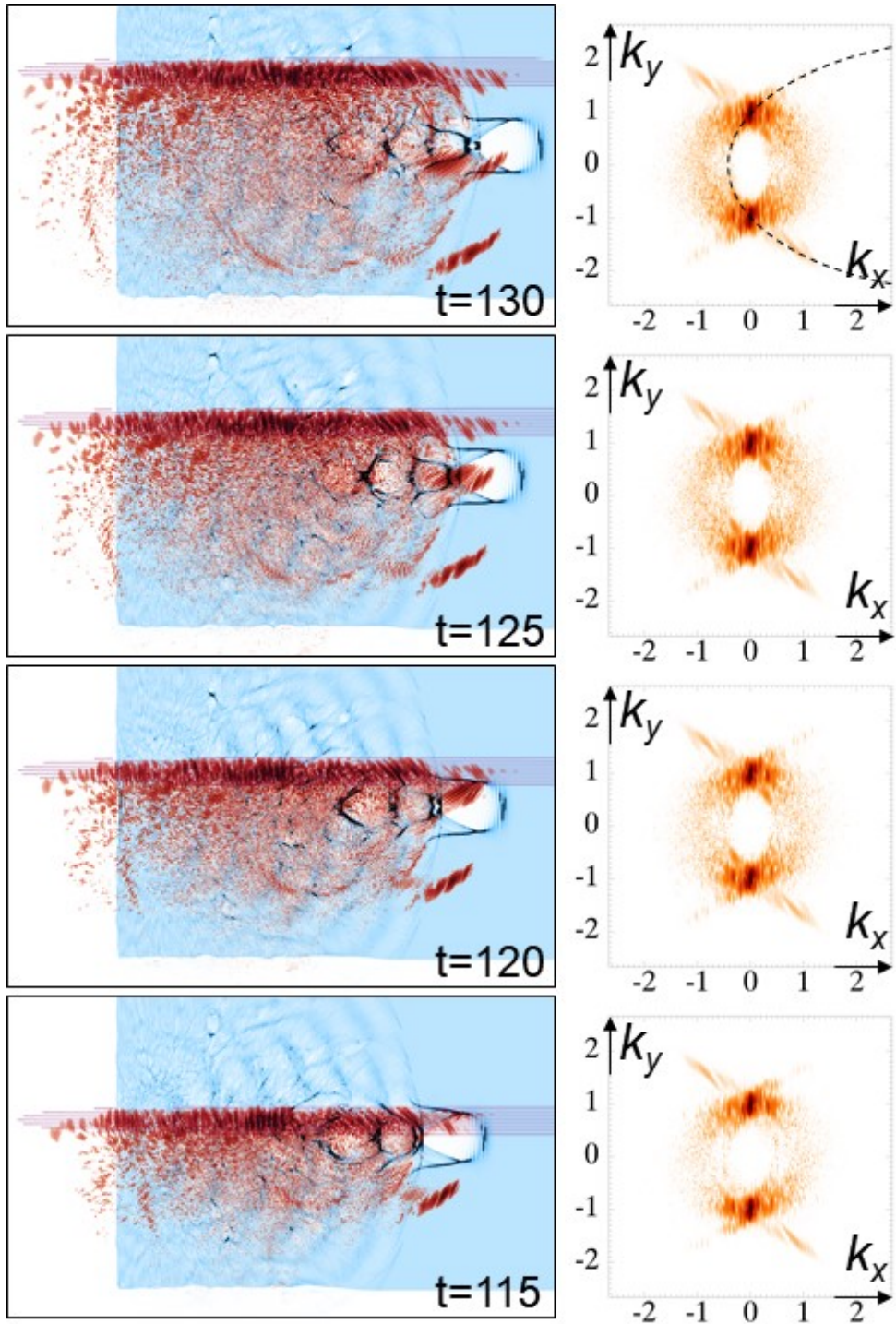

Fig. 6. The continuation of the time sequence of Fig. 5.

In Figs. 5 and 6, one can see the appearance of characteristic "protrusions" corresponding to high-frequency waves, which correlate with the onset of diffraction at the cusps. The "protrusions" are the portions of an ellipse given by the formula $\omega/c = (k_x^2+k_y^2)^{1/2} = \omega_0/(1-\beta \cos\alpha)$, where $\beta$ is the velocity of the cusp normalized by $c$, $\tan\alpha = k_y/k_x$. In the direction of the cusp motion, the diffracted wave frequency is upshifted due to the double Doppler effect, in the opposite direction it is downshifted. We note that even though the frequency gets upshifted and downshifted in different directions, the diffracted wave fronts are spherical. The diffraction on cusps can be considered both as transmission in the direction of the probe and as reflection in different directions. The reflection from relativistic cusps will be described soon elsewhere.

## Conclusion

Using analytical estimates and 2D PIC simulations we show that the Schlieren imaging of relativistically moving singularities in laser plasma is feasible. We derive the required duration of the optical probe pulse ensuring discernibility of different singularities in the images.

## Acknowledgements

We acknowledge financial support from the QST Director Fund 創成的研究 20.